\documentclass[prl,twocolumn,amsmath,amssymb,floatfix,letter]{revtex4}
\usepackage{graphicx}
\usepackage{bm}
\usepackage{amsmath,amssymb}
\usepackage{dcolumn}
\usepackage{bm}
\usepackage{epsfig}
\usepackage{longtable}
\usepackage{subfigure}
\usepackage{multirow}

\begin{document}

\title{{\rm\small\hfill }\\
    Beyond the Random Phase Approximation for the Electron Correlation Energy: The Importance of Single Excitations}
\author{Xinguo Ren}
\affiliation{Fritz-Haber-Institut der Max-Planck-Gesellschaft,
Faradayweg 4-6, 14195, Berlin, Germany}
\author{Alexandre Tkatchenko}
\affiliation{Fritz-Haber-Institut der Max-Planck-Gesellschaft,
Faradayweg 4-6, 14195, Berlin, Germany}
\author{Patrick Rinke}
\affiliation{Fritz-Haber-Institut der Max-Planck-Gesellschaft,
Faradayweg 4-6, 14195, Berlin, Germany}
\author{Matthias Scheffler}
\affiliation{Fritz-Haber-Institut der Max-Planck-Gesellschaft,
Faradayweg 4-6, 14195, Berlin, Germany}

\begin{abstract}
The random phase approximation (RPA) for the electron correlation energy, combined with the exact-exchange 
(EX) energy, represents the state-of-the-art exchange-correlation functional within density-functional theory (DFT). However, the standard RPA practice -- evaluating both the EX and the RPA correlation energies using Kohn-Sham (KS) orbitals from local or semi-local exchange-correlation functionals -- leads to a systematic underbinding of molecules and solids. Here we demonstrate that this behavior can be corrected by adding a ``single excitation" (SE) contribution, so far not included in the standard RPA scheme. A similar improvement can also be achieved by replacing the non-self-consistent EX total energy by the corresponding self-consistent Hartree-Fock total energy, while retaining the RPA correlation energy evaluated using KS orbitals. Both schemes achieve chemical accuracy for a standard benchmark set of non-covalent intermolecular interactions.
\end{abstract}

\maketitle

In the quest for finding an ``optimal" electronic structure method, that combines accuracy and tractability with transferability across different chemical environments and dimensionalities (e.g. molecules, wires/tubes, surfaces, solids), the treatment of exchange and correlation in terms of ``exact-exchange plus correlation in the random-phase approximation (EX+cRPA)" \cite{Bohm/Pines:1953,Gell-Mann/Brueckner:1957} offers 
a promising avenue
\cite{Furche:2001,Fuchs/Gonze:2002,Furche:2008,Scuseria/Henderson/Sorensen:2008,Janesko/Henderson/Scuseria:2009,Toulouse/etal:2009,Paier/etal:2010,Marini/Gonzalez/Rubio:2006,Harl/Kresse:2009,Lu/Li/Rocca/Galli:2009,Rohlfing/Bredow:2008,Ren/Rinke/Scheffler:2009,Schimka/etal:2010,Zhu/etal:2010}. In this approach, part of the exact-exchange (EX) energy cancels exactly the spurious self-interaction error present in the Hartree energy. The RPA correlation (cRPA) energy is fully non-local, whereby long-range van der Waals (vdW) interactions are included automatically and accurately \cite{Dobson:1994}). Moreover, dynamical electronic screening is taken into account by summing up a sequence of ``ring" diagrams to infinite order, making EX+cRPA applicable to small-gap or metallic systems where, for example, Hartree-Fock (HF) plus 2nd-order M{\o}ller-Plesset (MP2) perturbation theory \cite{Moller/Plesset:1934} breaks down. 

The concept of cRPA dates back to the many-body treatment of the uniform electron gas in the 1950's 
\cite{Bohm/Pines:1953,Gell-Mann/Brueckner:1957},
and was later formulated \cite{Langreth/Perdew:1977} within the context of density-functional theory 
(DFT) \cite{Kohn/Sham:1965}. 
Recent years have witnessed a revived interest in EX+cRPA  and its variants  in quantum chemistry \cite{Furche:2001,Fuchs/Gonze:2002,Furche:2008, Scuseria/Henderson/Sorensen:2008,Janesko/Henderson/Scuseria:2009,Toulouse/etal:2009,Paier/etal:2010}, solid state physics \cite{Marini/Gonzalez/Rubio:2006,Harl/Kresse:2009,Lu/Li/Rocca/Galli:2009}, and surface science  \cite{Rohlfing/Bredow:2008,Ren/Rinke/Scheffler:2009,Schimka/etal:2010}.
Within the framework of Kohn-Sham (KS) DFT, EX+cRPA embodies an orbital-dependent functional that can in principle be solved self-consistently via the optimized effective potential approach \cite{Kuemmel/Kronik:2008}. 
This is however numerically very demanding, and practical EX+cRPA calculations are commonly performed in a post-processing fashion, where single-particle orbitals from a self-consistent DFT calculation in the local-density approximation (LDA), generalized gradient approximations (GGAs), or alike, are used to evaluate both the EX and cRPA terms.  Alternatively, one can formulate cRPA in terms of many-body perturbation theory (MBPT) based on a Hartree-Fock (HF) reference.

Throughout this Letter we will adopt the following nomenclature: $E^F$@SC is the total energy of the functional $F$, evaluated with the orbitals of a self-consistent (SC) scheme, e.g., HF, or the Perdew-Burke-Ernzerhof (PBE)  \cite{Perdew/Burke/Ernzerhof:1996} GGA. The corresponding theoretical scheme is then labelled as $F$@SC. We also use the letter ``x" or ``c" in front of $F$ or as a subscript of $E^F$ to refer to the exchange or correlation part of the scheme explicitly.   The functional $F$ can be exact-exchange (EX), or additionally contain the RPA correlation (EX+cRPA), etc.  For instance, $E^\text{EX}$@HF is the self-consistent Hartree-Fock energy, whereas the conventional RPA scheme based on PBE orbitals is referred to as (EX+cRPA)@PBE.

The original (EX+cRPA)@PBE and (EX+cRPA)@HF schemes both exhibit systematic underbinding for a large variety of systems, including covalent molecules \cite{Furche:2001}, weakly bonded molecules  \cite{Janesko/Henderson/Scuseria:2009,Toulouse/etal:2009},  solids \cite{Harl/Kresse:2009}, and molecules adsorbed on surfaces \cite{Rohlfing/Bredow:2008,Ren/Rinke/Scheffler:2009,Schimka/etal:2010}.
Several attempts have been made to improve the accuracy of EX+cRPA. The earliest is the so-called RPA+ scheme \cite{Yan/Perdew/Kurth:2000} where a local correction at the LDA/GGA level is added to cRPA. More recent attempts add second-order screened exchange (SOSEX) \cite{Grueneis/etal:2009,Paier/etal:2010}) to make the entire approach self-correlation free, or invoke cRPA in a range-separated framework where only the long-range part of cRPA is incorporated \cite{Janesko/Henderson/Scuseria:2009,Toulouse/etal:2009}. Among these, RPA+ improves total correlation energies considerably \cite{Jiang/Engel:2007}, but not binding energies \cite{Furche:2001}. The SOSEX correction performs well \cite{Grueneis/etal:2009,Paier/etal:2010} with considerable additional numerical effort. Range-separated RPA schemes also improve upon the standard EX+cRPA scheme \cite{Janesko/Henderson/Scuseria:2009,Toulouse/etal:2009,Zhu/etal:2010}, however, at the price of introducing empirical parameters in the approach. 

In this Letter, we offer a new perspective, based on MBPT, for going beyond cRPA, and show that a simple modification of the standard EX+cRPA scheme leads to a significant accuracy increase for molecular binding energies. We first illustrate our key idea using the example of Ar$_2$. 
\begin{figure}
  \includegraphics[width=0.45\textwidth,clip]{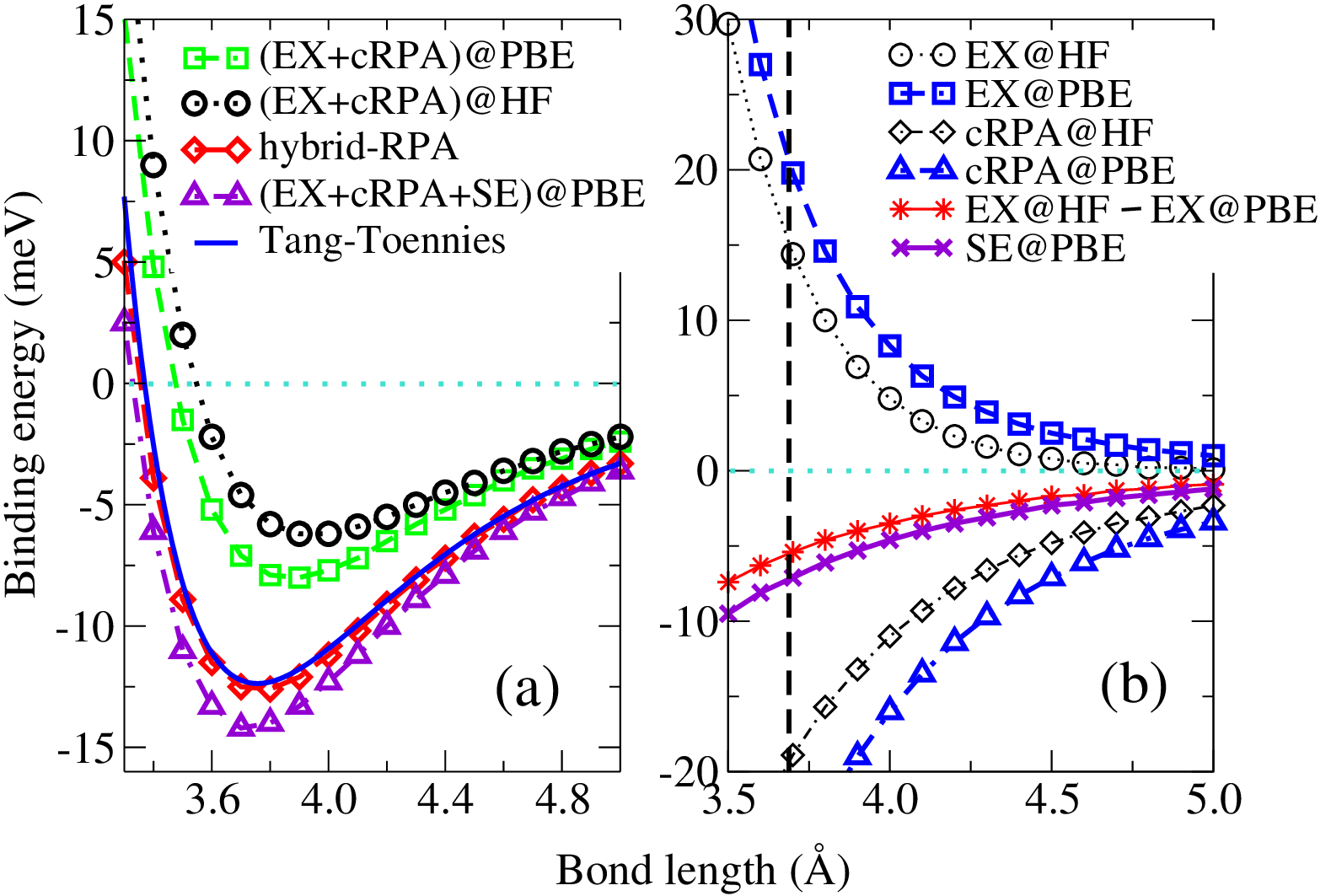}
   \caption{(Color online) Panel (a): Binding energy curve for Ar$_2$ computed with four 
           RPA-based approaches, in comparison to the accurate reference curve by Tang and 
           Toennies \cite{Tang/Toennies:2003}.  Panel (b): Decomposition of the 
           (EX+cRPA)@HF ((EX+cRPA)@PBE) binding energy of Ar$_2$
           into individual contributions: EX@HF (EX@PBE) and 
           cRPA@HF (cRPA@PBE). The difference
           between EX@HF and EX@PBE, and the SE@PBE
           term are also plotted. The vertical dashed line marks the equilibrium distance. 
           Calculations are done using FHI-aims \cite{Blum/etal:2009,Ren/inpreparation} and 
           Dunning's aug-cc-pV6Z basis \cite{Dunning:1989}. The basis set superposition error 
           (BSSE) is corrected here and in the following.
           }
  \label{Fig:Ar2_BE}
\end{figure}
The (EX+cRPA)@PBE and (EX+cRPA)@HF binding energy curves for Ar$_2$ are plotted in Fig.~\ref{Fig:Ar2_BE}(a). Both schemes show a significant underbinding behavior compared to the reference curve modeled by Tang and Toennies \cite{Tang/Toennies:2003} based on experimental data. To gain more insight into the origin of the underbinding, the EX+cRPA binding energies are decomposed into two contributions in Fig.~\ref{Fig:Ar2_BE}(b): the exchange-only part and the remaining cRPA part. Inspection of the individual components reveals that $E^\text{cRPA}_\text{c}$@HF is (much) more repulsive than  $E^\text{cRPA}_\text{c}$@PBE, whereas at the EX level $E^\text{EX}$@PBE is (much) more repulsive than  $E^\text{EX}$@HF. The fact that  $E^\text{cRPA}_\text{c}$@PBE is more attractive than $E^\text{cRPA}_\text{c}$@HF is easy to rationalize by inspecting the corresponding frequency-dependent polarizabilities. Extensive benchmark calculations for 1225 molecular pairs \cite{Tkatchenko/Scheffler:2009} show that asymptotic $C_6$ dispersion coefficients derived from 
$E^\text{cRPA}_\text{c}$@HF are systematically too small by approximately $40\%$ \cite{Ren/inpreparation}, while this error is only $\sim 10\%$ for $E^\text{cRPA}_\text{c}$@PBE. Adding $\Delta$vdW corrections in an attempt to reduce the remaining error in cRPA@PBE \cite{Tkatchenko/etal:2009} only leads to minor changes in the binding energy at the equilibrium distance. What is more striking, however, is the considerable difference in binding energies at the EX level ---- $E^\text{HF}$@HF$-E^\text{EX}$@PBE (plotted also in Fig.~\ref{Fig:Ar2_BE}(b) (red stars)). It amounts to $\sim$6 meV at the equilibrium distance and is thus close to the deviation of the (EX+cRPA)@PBE binding energy from the reference value. 

From the viewpoint of Rayleigh-Schr{\"o}dinger perturbation theory (RSPT), $E^\text{EX}$@HF and 
$E^\text{EX}$@PBE correspond to the sum of the zeroth and first-order terms in the perturbative expansions  
based on HF and PBE reference state respectively \cite{Szabo/Ostlund:1989}. The difference between $E^\text{EX}$@HF and $E^\text{EX}$@PBE must therefore be compensated by higher-order terms in the perturbation series since the final result should be independent of the reference state, if all terms were summed up. The next term in the series is the 2nd-order correlation energy $E_\text{c}^\text{(2)}$, to which only single and double excitation configurations contribute. Here we particularly examine the contribution of single excitations (SE) to $E_\text{c}^\text{(2)}$, which can be expressed \cite{Szabo/Ostlund:1989} as
   \begin{equation}
    E^\text{SE}_\text{c} = \sum_{i}^\text{occ}\sum_{a}^\text{unocc} 
    \frac{|\langle\psi_i|\hat{f}|\psi_a\rangle|^2}{\epsilon_i-\epsilon_a}. 
    \label{Eq:sec}
   \end{equation}
Here $\psi_i$ and $\epsilon_i$ are the single particle orbitals and orbital energies of the reference state, and $\hat{f}$ is the single-particle HF Hamiltonian -- the Fock operator. A more detailed derivation of Eq.~(\ref{Eq:sec}) is given in the supplementary material (where we simply follow the RSPT instead of the
G{\"o}rling-Levy PT \cite{Goerling/Levy:1993}). As a consequence of the Brillouin theorem \cite{Szabo/Ostlund:1989}, $E^\text{SE}_\text{c}$ trivially vanishes for HF orbitals, but is in general non-zero for KS orbitals. The contribution of $E^\text{SE}_\text{c}$ evaluated with PBE orbitals (referred to as SE@PBE) to the binding energy of Ar$_2$ is  plotted in Fig.~\ref{Fig:Ar2_BE}(b) (violet crosses). It amounts to $50\%$ of the binding energy at the equilibrium distance, and is close in magnitude to the contribution from $E^\text{EX}$@HF$-E^\text{EX}$@PBE, and to the amount of underbinding in the original (EX+cRPA)@PBE scheme. We therefore propose
a new scheme by adding $E^\text{SE}_\text{c}$ to $E^\text{EX+cRPA}$ (subsequently referred to as EX+cRPA+SE).
In Fig.~\ref{Fig:Ar2_BE}(a) the resultant (EX+cRPA+SE)@PBE binding energy curve is also plotted, which
improves considerably over the (EX+cRPA)@PBE results, and is in close agreement with the Tang-Toennies 
reference curve.  

It appears that the quantitative agreement between $E^\text{SE}_\text{c}$ defined in Eq.~(\ref{Eq:sec}) and $E^\text{EX}$@HF$-E^\text{EX}$@PBE is a general feature. We found for a set of 50 atoms and molecules that the agreement  typically ranges between $70\%$ and $100\%$, suggesting that replacing $E^\text{EX}$@PBE by $E^\text{EX}$@HF is an effective way to account for the SE contributions. This leads to a ``hybrid-RPA" scheme, whose total energy is given by
\begin{equation}
  E^\text{hybrid-RPA} = E^\text{EX}@\text{HF} + E_\text{c}^\text{cRPA}@\text{PBE},
  \label{Eq:rpah}
\end{equation}
as an alternative to boost the accuracy of RPA. Fig.~\ref{Fig:Ar2_BE}(a) shows that the resultant binding energy curve is in almost perfect agreement with the reference curve.

\begin{figure}
  \includegraphics[width=0.4\textwidth,clip]{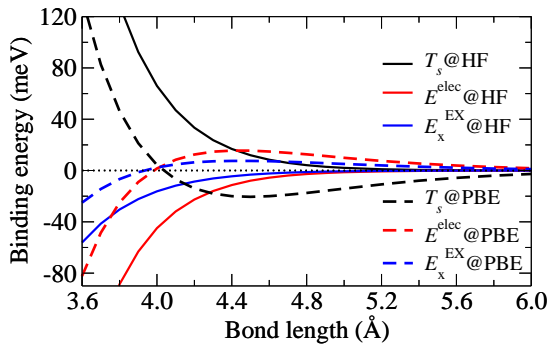}
   \caption{(Color online) Decomposition of the $E^\text{EX}$@HF and $E^\text{EX}$@PBE binding energies for Ar$_2$  into their kinetic, electrostatic, and EX components. }
  \label{Fig:Ar2_component}
\end{figure}

At this point, it is illustrative to take a closer look at the individual contributions to $E^\text{EX}$@HF$-E^\text{EX}$@PBE. In Fig. \ref{Fig:Ar2_component} we further decompose the EX@HF and EX@PBE  binding energies into their kinetic ($T_\text{s}$), electrostatic ($E^\text{elec}$, external potential energy and Hartree energy combined), and EX components ($E^\text{EX}_\text{x}$) for Ar$_2$.
All three energy components behave quite differently for HF and PBE orbitals. The HF kinetic energy is purely repulsive, whereas the PBE one exhibits spurious attraction at intermediate and large distances. The HF electrostatic and exact-exchange energies, on the other hand, are purely attractive and decay to zero from below, while the corresponding PBE ones become repulsive in the intermediate range and decay to zero from above at large distances. Since the PBE orbitals are less localized than their HF counterparts all three energy components decay much slower in PBE than
in HF. The overall effect is that $E^\text{EX}$@PBE becomes significantly more repulsive than $E^\text{EX}$@HF, resulting in the underbinding behavior of (EX+cRPA)@PBE. The more physical behavior of EX@HF than EX@PBE at
 the EX level provides a sound basis for the systematic improvement from (EX+cRPA)@PBE to hybrid-RPA.

\begin{figure}
\includegraphics[width=0.45\textwidth]{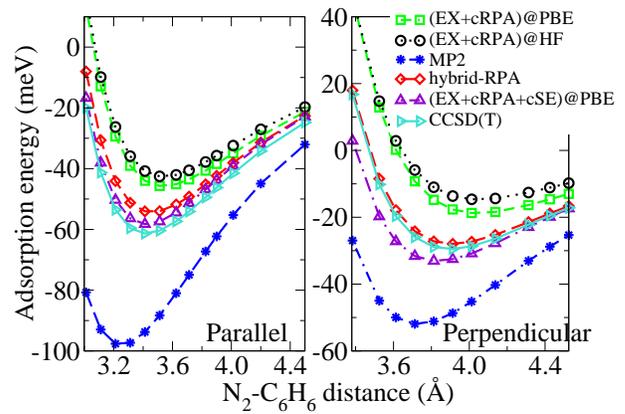}
\caption{
	(Color online) Binding energies of the parallel and perpendicular configuration of
	N$_2$@benzene as a function of the N$_2$-C$_6$H$_6$ center-of-mass distance, 
	calculated by different RPA-based approaches as well as MP2 compared to reference
	CCSD(T) calculations from Ref.~[\onlinecite{Tkatchenko/etal:2010}]. 
}
\label{Fig:N2onBenzene}
\end{figure}

Indeed, the exceptional performance of the hybrid-RPA and (EX+cRPA+SE)@PBE schemes for rare-gas dimers carries over to many other molecular systems. As a second example we show results for the N$_2$ molecule adsorbed on benzene (N$_2$@benzene), which is an important model system for studying molecular adsorption on graphene and graphite surfaces~\cite{Tkatchenko/etal:2010}. We consider two possible configurations: N$_2$ placed parallel or perpendicular to the benzene plane. A successful theoretical approach for this system must be able to describe the delicate balance between electrostatic and dispersion interactions. We use FHI-aims \cite{Blum/etal:2009,Ren/inpreparation} numeric atom-centered orbital basis ($6s5p4d3f2g$ for C, O, N, and $5s3p2d1f$ for H ) augmented with gaussian diffuse functions from aug-cc-pV5Z to achieve convergence of the binding energy to within 1 meV. The results shown in Fig.~\ref{Fig:N2onBenzene} are very similar to the rare-gas dimers: (EX+cRPA)@HF and 
 (EX+cRPA)@PBE underbind significantly at the equilibrium distance, while hybrid-RPA and (EX+cRPA+SE)@PBE bring the binding energy into much closer agreement with the reference curve computed with the coupled cluster method including single, double and perturbative triple excitations (CCSD(T)) \cite{Tkatchenko/etal:2010}. In contrast, the traditional MP2 
method vastly overbinds the system.

\begin{figure}
\includegraphics[width=0.45\textwidth]{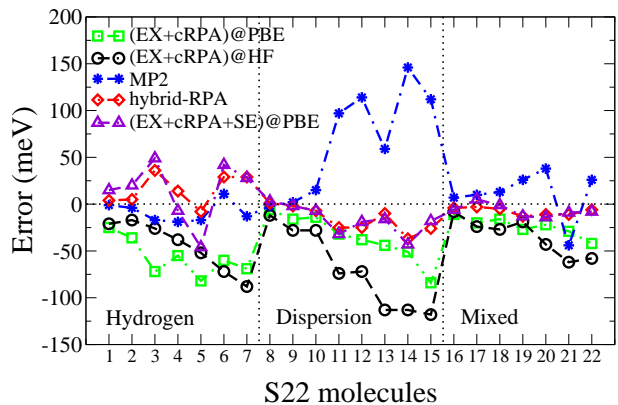}
\caption{
(Color online) Deviation from the CCSD(T) reference values \cite{Takatani/etal:2010} for the binding energies of the S22 database \cite{Jurecka/etal:2006} for RPA-based approaches as well as MP2. Positive errors correspond to overbinding and negative ones to underbinding.
}\label{Fig:S22}
\end{figure}

Finally we examine the performance of hybrid-RPA and (EX+cRPA+SE)@PBE for the S22 database of Jure\v{c}ka \textit{et al.} \cite{Jurecka/etal:2006}, which represents a balanced benchmark set for non-covalent interactions. The molecular dimers in this database can be divided into three groups of different bonding types: hydrogen-bonded, dispersion-bonded, and mixed complexes. We note that RPA in a range-separated framework has been applied to the S22 database very recently \cite{Zhu/etal:2010}. In Fig.~\ref{Fig:S22} we plot the deviation from the CCSD(T) reference values \cite{Takatani/etal:2010} for the binding energies in the S22 database \cite{Jurecka/etal:2006} for four RPA-based approaches and MP2. The basis set type and quality is the same as for N$_2$@benzene. A detailed error analysis is presented in Table~\ref{Tab:S22}.

\begin{table}
\caption{
Mean absolute error (in meV) and mean absolute percentage error (in parenthesis) of different 
RPA-based approaches for the S22 database \cite{Jurecka/etal:2006}. CCSD(T) extrapolated to the complete basis set limit ~\cite{Takatani/etal:2010} is taken as reference. 
}
\centering
\begin{tabular}{lccc}
\hline
\mbox{} & \mbox{H-bond} & \mbox{Dispersion} & \mbox{Mixed} \\ \hline 
\hline
(EX+cRPA)@HF           & 45 (~8.5\%)  & 70 (43.9\%) & 34 (20.9\%) \\ 
(EX+cRPA)@PBE          & 57 (11.2\%) & 36 (21.8\%) & 24 (15.0\%) \\ 
(EX+cRPA+SE)@PBE      &  30 (~6.0\%)  & 18 (12.0\%) & 8 (~5.5\%) \\ 
hybrid-RPA        & 18 (~3.0\%)  & 17 (10.0\%) & 8 (~5.1\%)  \\
\hline
\end{tabular}
\centering
\label{Tab:S22}
\end{table}
We observe that the standard (EX+cRPA)@PBE scheme systematically underbinds all complexes. (EX+cRPA)@HF performs even worse for dispersion and mixed bonding, but better for hydrogen bonding. The latter case can be explained by the fact that the better performance of EX@HF dominates over the bad performance of cRPA@HF for hydrogen bonded systems. Again hybrid-RPA and (EX+cRPA+SE)@PBE correct the underbinding behavior of the standard EX+cRPA scheme, and improve the accuracy considerably. The hybrid-RPA scheme yields a mean absolute error (MAE) of 14 meV. The performance of (EX+cRPA+SE)@PBE is very similar to hybrid-RPA for dispersion and mixed bonding, albeit somewhat worse for  hydrogen bonding. However, the mean absolute percentage error for hydrogen bonding (6$\%$) is still quite small. 
The accuracies achieved here compare favorably to the recently developed vdW functional (vdW-DF) \cite{Dion/etal:2004}, 
where the MAE for the PBE-based vdW-DF results for S22 \cite{Gulans/Puska/Nieminen:2009} is 54 meV.
We also note that for covalent molecules the accuracies in the atomization energies 
are improved considerably by the two schemes. For instance, the MAE of the atomization energies of the G2-I set is 
reduced from 10.5 kcal/mol to 6.2 kcal/mol by (EX+cRPA+SE)@PBE and 6.3 kcal/mol by hybrid-RPA.

To summarize, we have unraveled the origin of the underbinding that plagues the standard (EX+cRPA)@PBE scheme, which is mostly due to the too-repulsive nature of $E^\text{EX}$@PBE rather than the (slightly) 
underestimation of the long-range dispersion force by $E^\text{cRPA}_\text{c}$@PBE. This problem can be largely solved either by replacing $E^\text{EX}$@PBE by the self-consistent HF energy $E^\text{EX}$@HF, or by adding a SE correction to the standard (EX+cRPA)@PBE approach. Particularly 
 (EX+cRPA+SE)@PBE is a well-defined parameter-free scheme in which the SE term does not add any significant computational cost to the approach.  In addition, the SE correction is compatible with other beyond-RPA schemes like RPA+ or SOSEX. 
We also like to emphasize that in both schemes the cRPA evaluated with KS orbitals 
is retained, which is essential for producing quantitatively correct asymptotics for vdW bonded systems.
Despite its success for describing vdW and covalently bonded molecules, one obvious deficiency of the 
2nd-order SE as given by Eq.~(\ref{Eq:sec}), however, is that it is not well-behaved for 
systems with vanishing gaps. In such cases, we propose to  ``renormalize" the SE 
contribution  via a resummation of a geometrical series of higher-order diagrams involving single 
excitations (in the spirit of cRPA).  This leads to additional terms in the denorminator of 
Eq.~(\ref{Eq:sec}) which prevent the possible divergence even when the KS gap closes.  
A brief derivation of this renormalized SE (RSE) scheme is presented in the supplementary material.
Further details and benckmark calculations will be published elsewhere \cite{Ren/inpreparation}.




\end{document}



\section{Derivation of the single excitation contribution to the 2nd order correlation energy}
In this section we derive Eq. (1) that is presented in the main part of this Letter -- the single excitation contribution to the 2nd-order correlation energy -- from Rayleigh-Schr\"odinger perturbation theory (RSPT).   The interacting $N$-electron system at hand is governed  by the Hamiltonian 
 \begin{equation}
   \hat{H} = \sum_{j=1}^{N}\left[-\frac{1}{2}\nabla^2_j + \hat{v}_\text{ext} (\bfr_j) \right]  + 
     \sum_{j<k}^{N} \frac{1}{|\bfr_j -\bfr_k|}, \nonumber
 \end{equation}
where  $\hat{v}_\text{ext}(\bfr)$ is a local, multiplicative external potential. In RSPT, $\hat{H}$ is partitioned into a non-interacting mean-field  Hamiltonian $\hat{H}^0$ and an interacting perturbation $\hat{H}'$,
 \begin{eqnarray} 
   \hat{H} & =& \hat{H}^0 + \hat{H}'    \nonumber  \\
   \hat{H}^0 &=& \sum_{j=1}^{N} \hat{h}^0 (j) = \sum_{j=1}^{N} \left[-\frac{1}{2}\nabla^2_j + \hat{v}_\text{ext} (\bfr_j) + 
         \hat{v}^\text{MF}_j \right] \nonumber \\
   \hat{H}' &=& \sum_{j<k}^{N} \frac{1}{|\bfr_j -\bfr_k|} - \sum_{j=1}^N \hat{v}^\text{MF}_j. \nonumber
 \end{eqnarray}
Here $\hat{v}^\text{MF}$ is some mean-field-type single-particle potential, which can be non-local, as 
in the case of Hartree-Fock (HF) theory, or local, as in the case of Kohn-Sham (KS) theory.

Suppose the solution of the single-particle Hamiltonian $\hat{h}^0$ is known 
  \begin{equation}
    \hat{h}^0 |\psi_p\rangle = \epsilon_p |\psi_p\rangle,
   \label{Eq:SP_eigeneq}
  \end{equation}
then the solution of the non-interacting many-body Hamiltonian $H^0$ follows 
 \begin{equation}
   \hat{H}^0 |\Phi_n \rangle = E^{(0)}_n |\Phi_n \rangle  \nonumber .
 \end{equation}
The $|\Phi_n \rangle$ are Slater determinants formed from $N$ of the spin orbitals $|p\rangle = |\psi_p\rangle$ determined in Eq.~(\ref{Eq:SP_eigeneq}).  These Slater determinants can be distinguished according to their excitation level: the ground-state configuration $|\Phi_0\rangle$, singly excited  configurations $|\Phi_i^a\rangle$, doubly excited configurations $|\Phi_{ij}^{ab}\rangle$, etc., where $i,j, \dots$ denotes occupied orbitals and $a,b,\dots$ unoccupied ones. Following standard perturbation theory, the  single-excitation (SE) contribution to the 2nd-order correlation energy is given by
 \begin{eqnarray}
  E^\text{SE}_c& =& \sum_{i}^\text{occ}\sum_a^\text{unocc}\frac{|\langle\Phi_0|\hat{H}'|\Phi_i^a\rangle|^2}{E^{(0)}_0 - E^{(0)}_{i,a}} \nonumber \\
               & =&\sum_{i}^\text{occ}\sum_a^\text{unocc}\frac{|\langle\Phi_0|\sum_{j<k}^{N} \frac{1}{|\bfr_j -\bfr_k|} - \sum_{j=1}^N \hat{v}^\text{MF}_j|\Phi_i^a\rangle|^2}{\epsilon_i - \epsilon_a} \nonumber \\
  \label{Eq:SE_expression}
 \end{eqnarray}
where we have used the fact $E^{(0)}_0 - E^{(0)}_{i,a} = \epsilon_i - \epsilon_a$.

To proceed, the numerator of Eq.~(\ref{Eq:SE_expression}) needs to be evaluated. This can be most easily done using  second-quantization formulation
  \begin{eqnarray}
    \sum_{j<k}^{N} \frac{1}{|\bfr_j -\bfr_k|} & \rightarrow & \frac{1}{2}\sum_{pqrs} \langle pq|rs \rangle c_p^{\dagger} 
             c_q^{\dagger}c_sc_r, \nonumber \\
    \sum_{i=j}^N \hat{v}^\text{MF}_j & \rightarrow & \sum_{pq} \langle p|\hat{v}^\text{MF}|q \rangle c_p^{\dagger} c_q, \nonumber
  \end{eqnarray}
where $p,q,r,s$ are arbitrary spin-orbitals from Eq.~(\ref{Eq:SP_eigeneq}), $c_p^{\dagger}$ and $c_q$, etc.
are the electron creation and annihilation operators, and $ \langle pq|rs \rangle$ the two-electron Coulomb integrals
  \begin{equation}
      \langle pq|rs \rangle = \int d\bfr d\bfrp \frac{\psi_p^\ast(\bfr)\psi_r(\bfr)\psi_q^\ast(\bfrp)\psi_s(\bfrp)}
            {|\bfr-\bfrp|}. \nonumber
  \end{equation}
The expectation value of the two-particle Coulomb operator between the ground-state configuration $\Phi_0$ and the single excitation $\Phi_i^a$ is given by
  \begin{eqnarray}
    \displaystyle
   \langle\Phi_0| \frac{1}{2}\sum_{pqrs} \langle pq|rs \rangle c_p^{\dagger}c_q^{\dagger}c_sc_r |\Phi_i^a\rangle & =&
    \sum_p^\text{occ} \left[ \langle ip|ap\rangle - \langle ip|pa\rangle \right] \nonumber \\
      & =& \langle \psi_i |\hat{v}^\text{HF}|\psi_a\rangle
   \label{Eq:TP_operator}
 \end{eqnarray}
where $v^\text{HF}$ is the HF single-particle potential. 

The expectation value of the mean-field single-particle operator $\hat{v}^\text{MF}$, on the other hand, is given by
 \begin{equation}
    \langle\Phi_0|  \sum_{pq} \langle p|\hat{v}^\text{MF}|q \rangle c_p^{\dagger} c_q | \Phi_i^a\rangle 
    = \langle \psi_i |\hat{v}^\text{MF}|\psi_a\rangle 
   \label{Eq:SP_operator}
 \end{equation}
Combining Eqs.~(\ref{Eq:SE_expression}), (\ref{Eq:TP_operator}), and (\ref{Eq:SP_operator}), one gets
\begin{eqnarray}
  E_{c}^\text{SE} &=& 
   \sum_{i}^\text{occ}\sum_a^\text{unocc}
           \frac{|\langle \psi_i |\hat{v}^\text{HF}-\hat{v}^\text{MF}|\psi_a\rangle|^2}{\epsilon_i - \epsilon_a} 
          \nonumber \\
   &=& \sum_{i}^\text{occ}\sum_a^\text{unocc} \frac{|\Delta v_{ia}|^2}{\epsilon_i - \epsilon_a},
  \label{Eq:2nd_cSE}
\end{eqnarray}
where $\Delta v_{ia}$ is the matrix element of the difference between the HF potential $\hat{v}^\text{HF}$ 
and the single-particle mean-field potential $\hat{v}^\text{MF}$ in question.

Observing that the $\psi$'s are eigenstates of $\hat{h}^0=-\frac{1}{2}\nabla^2 + v_\text{ext}+v^\text{MF}$, and hence all non-diagonal elements $\langle\psi_i|\hat{h}^0|\psi_a\rangle$ are zero, one can alternatively express 
Eq.~(\ref{Eq:2nd_cSE}) as
\begin{eqnarray}
  E_{c}^\text{SE} &=& 
    \sum_{i}^\text{occ}\sum_a^\text{unocc}\frac{|\langle \psi_i |-\frac{1}{2}\nabla^2 + \hat{v}_\text{ext} + 
       \hat{v}^\text{HF} |\psi_a\rangle|^2}{\epsilon_i - \epsilon_a} \nonumber \\
  &=& \sum_{i}^\text{occ}\sum_a^\text{unocc}
           \frac{|\langle \psi_i |\hat{f}|\psi_a\rangle|^2}{\epsilon_i - \epsilon_a} 
  \label{Eq:2nd_cSE_1}
\end{eqnarray}
where $\hat{f}$ is the single-particle HF Hamiltonian, or simply Fock operator. Thus Eq. (1) in the
main paper is derived.

For the HF reference state, i.e., when $\hat{v}^\text{MF} = \hat{v}^\text{HF}$, the $\psi$'s are eigenstates of the Fock operator, and hence Eq.~(\ref{Eq:2nd_cSE}) (or (\ref{Eq:2nd_cSE_1})) is zero. For any other reference state, e.g., the KS reference state, the  $\psi$'s are no longer eigenstates of the Fock operator, and Eq.~(\ref{Eq:2nd_cSE}) is in general not zero. This gives rise to a finite SE contribution to the second-order correlation energy. In the following we assume $\hat{v}^\text{MF} = \hat{v}^\text{KS}$. 

\section{Renormalized single excitation (RSE) correction scheme}

\begin{figure}
\includegraphics[width=0.46\textwidth,clip]{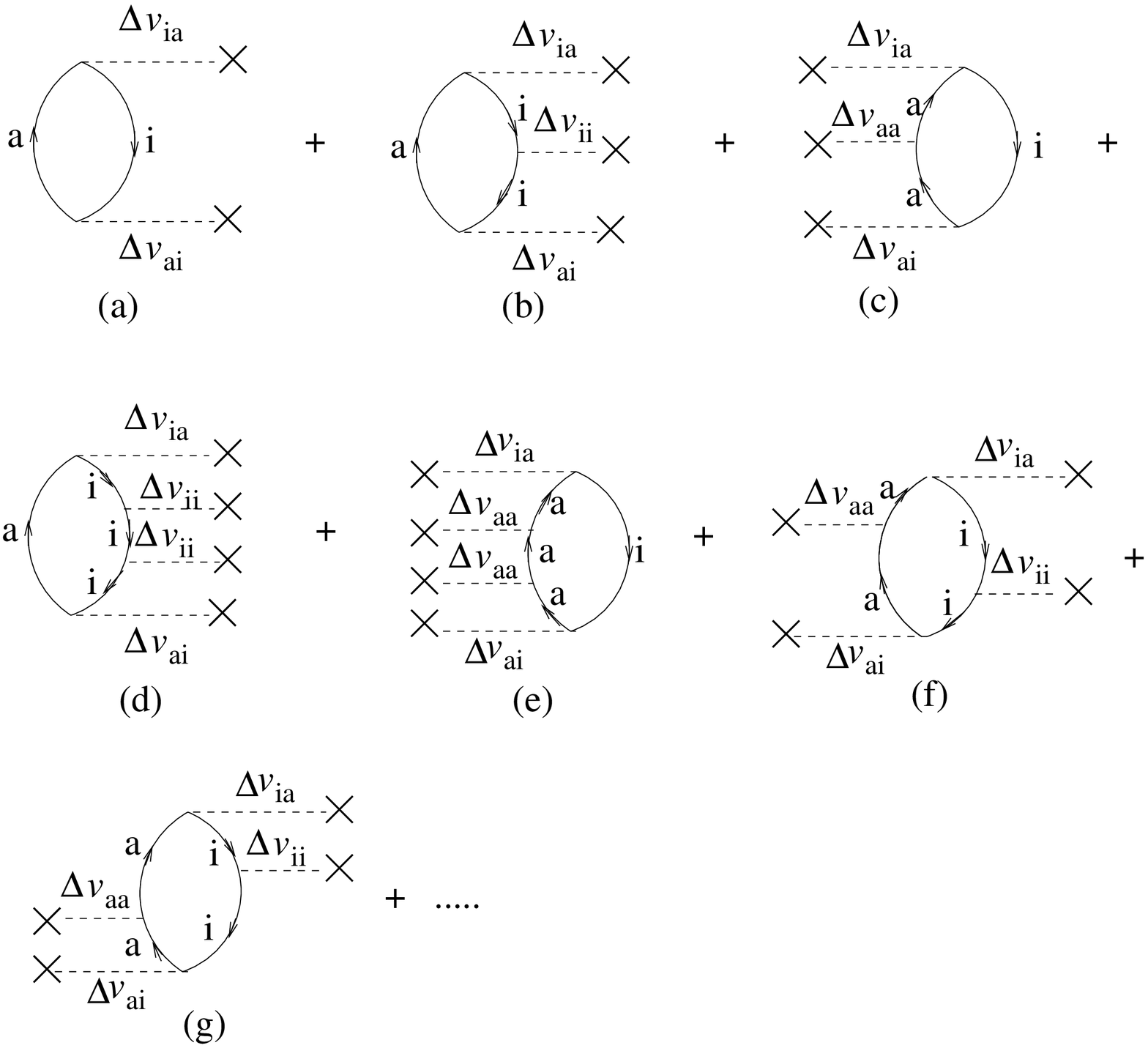}
\caption{Goldstone digrammatic representation of a selection of single excitation processes.
(a): second order; (b-c): third-order; (d-g): fourth-order. Here 
$v_{ii} =\langle \psi_i|\hat{v}^\text{HF}-\hat{v}^\text{KS}|\psi_i \rangle$ and 
$v_{aa} =\langle \psi_a|\hat{v}^\text{HF}-\hat{v}^\text{KS}|\psi_a \rangle$.}
\label{Goldstone_diag}
\end{figure}

The SE contribution as given by Eq.~(\ref{Eq:2nd_cSE}) corresponds to a correction to the correlation energy 
at 2nd-order, which can be represented by the Goldstone diagram \cite{Goldstone:1957} shown in 
Fig.~\ref{Goldstone_diag}(a).
Similar to 2nd-order M{\o}ller-Plesset perturbation theory, the 2nd-order SE is ill-behaved when the 
single-particle gap closes. An important lesson from diagrammatic perturbation theory to deal with 
problems of this kind is to resum higher order diagrams of the same type to infinite order. 
The random phase approximation itself is a good example of this. In case of SE, one can also 
identify a series of higher order diagrams up to infinite order, as illustrated 
in terms of Goldstone diagrams in Fig.~\ref{Goldstone_diag}.
Here we only pick  the ``diagonal" terms where a particle state $a$ or a hole state $i$ is, in the intermediate process, scattered into the same state by the perturbative potential
$\Delta \hat{v} = \hat{v}^\text{HF} - \hat{v}^\text{KS}$. These terms dominate over the ``non-diagonal"
ones \cite{magnidiag} and facilitate a simple mathematical treatment, because they form a geometrical series that can easily be summed up. Here we call
the summation of this series \textit{renormalized} single excitations (RSE).  Following
the textbook rule \cite{Szabo/Ostlund:1989} for evaluating the Goldstone diagrams, one obtains
\begin{eqnarray}
E^\text{RSE}_c &=& \sum_{i}^\text{occ}\sum_a^\text{unocc}  
                  \frac{|\Delta v_{ia}|^2}{\epsilon_i - \epsilon_a} +  \nonumber \\
            &&  \left[  \sum_{i}^\text{occ}\sum_a^\text{unocc}
                \frac{|\Delta v_{ia}|^2 (\Delta v_{aa} - \Delta v_{ii})}{(\epsilon_i - \epsilon_a)^2} \right] + \nonumber \\ 
            && \left[  \sum_{i}^\text{occ}\sum_a^\text{unocc}
                \frac{|\Delta v_{ia}|^2 (\Delta v_{aa} - \Delta v_{ii})^2}{(\epsilon_i - \epsilon_a)^3} \right] +  \nonumber \\
            &&  \cdots  \nonumber  \\
            &=& \sum_{i}^\text{occ}\sum_a^\text{unocc} \frac{|\Delta v_{ia}|^2}
                {\epsilon_i - \epsilon_a + \Delta v_{ii}-  \Delta v_{aa} }
        \label{Eq:rse}
\end{eqnarray}
The additonal term $\Delta v_{ii} - \Delta v_{aa}$ in the denominater of Eq.~(\ref{Eq:rse}), which appears to be 
negative definite in practical calculations, ensures that $E^\text{RSE}_c$ is finite and
prevents possible divergence problems even when the single-particle KS gap closes.
In this context, we note in passing that a similar partial resummation of the so-called Epstein-Nesbet 
series \cite{Szabo/Ostlund:1989} of diagrams to ``renormalize" the 2nd-order correlation 
energy has been performed by Jiang and Engel \cite{Jiang/Engel:2006} in the 
KS many-body perturbation theory.

\section{Acknowledgement}
XR thanks Rodney J. Bartlett, Georg Kresse, and Hong Jiang for helpful discussions.